\documentstyle[11pt]{article}

\if@twoside
\else
\fi
\begin{document}

\title{
                    \vspace{-4cm}
        \hfill \fbox{{\normalsize {\bf STPHY-TH/94-10}}}\\[1cm]
Spin in the path integral: anti-commuting versus commuting variables}

\author{
F G Scholtz, A N Theron and H B Geyer\\[5pt]
{\it Institute of Theoretical Physics}\\
{\it University of Stellenbosch,
7600 Stellenbosch,
South Africa}
}
\date{}

\maketitle

\begin{abstract}
We discuss the equivalence between the path integral representations
of spin dynamics for anti-commuting (Grassmann) and commuting
variables and establish a bosonization dictionary for both
generators of spin and single fermion operators.  The
content of this construction in terms of the representations of the
spin algebra is discussed in the path integral setting.  Finally it is
shown how a `free field realization' (Dyson mapping) can be
constructed in the path integral.
\end{abstract}

\baselineskip 17pt

The dynamics of a particle with spin coupled to an external magnetic field
$\alpha_i$ can be formulated in the path integral in terms of commuting
variables.  A transition matrix element is given by [1]
\begin{equation}
\int_{s_i}^{s_f}[ds]\exp(\int_{t_i}^{t_f}dt[i\lambda{\rm tr}(\dot
s^\dagger\sigma_3 s)+\lambda{\rm tr}(\sigma_3 s\alpha s^\dagger)])\,.\label{1a}
\end{equation}
Here $s$ is the spin half representation, $[ds]$ the invariant measure on the
coset U(1)$\setminus$SU(2), $\sigma_i$ the Pauli matrices,
$\lambda$ the spin of the particle, and $\alpha=\alpha_i\sigma_i$.

On the other hand the dynamics can also be formulated in terms of a path
integral over Grassmann variables $\eta$, a transition matrix element being
given by [2]
\begin{equation}
\int_{\eta_i}^{\eta_f}[d\eta]\exp(\int_{t_i}^{t_f}dt[i\eta\dot\eta+
i\epsilon_{ijk}\alpha_i\eta_j\eta_k])\,.\label{1b}
\end{equation}

One expects these two formulations to be equivalent.  Here we show that this is
indeed the case by transforming a slightly more general form of the path
integral representation (\ref{1b}) in terms of Grassmann variables into the
path integral (\ref{1a}) in terms of commuting variables.  This process is
popularly known as bosonization and plays a particularly important role in
many-body physics and field-theories in, mostly, 1+1 dimensions.  Indeed, the
model we are about to present is a toy model for non-abelian bosonization in
1+1 dimensions, leading to the Wess-Zumino-Witten model [3].  This model is
also important in the many-body physics setting, and a discussion on its
bosonization and the role it plays in the many-body problem can be found in
[4].

Naturally bosonization entails a set of rules that relate equivalent fermionic
and bosonic quantities.  This set of rules results immediately from our
procedure.  In particular it enables us to write down the bosonic equivalents
of the generators of the spin algebra as well as the bosonic equivalents of
single fermion operators in terms of vertex operators [5].

We first formulate our fermion model.  Consider the su(2)
representation [4]
\begin{equation}
J_+=\sum_{m>0}a^\dagger_m a^\dagger_{-m};\quad
J_-=\sum_{m>0}a_{-m} a_{m};\quad J_0=\sum_{m>0}(a^\dagger_m a_m+
a_{-m} a^\dagger_{-m})/2\,.\label{1}
\end{equation}
on fermion Fock space, with
$a^\dagger_m (a_m)\; m=\pm 1,\pm 2,\ldots,\pm\Omega$ fermion
creation
(annihilation) operators.  This representation is clearly
reducible and Fock space carries many equivalent and inequivalent irreps.
This is indeed our motivation for considering the specific model:
non-trivial tensor operators on Fock space can be constructed by
appropriate couplings of fermion operators and their bosonic
equivalents sought.
We focus especially on
single fermion operators $a^\dagger_m\,,a_{-m}$, which belong to a doublet, and
out of which all other tensor operators can be build.
Of particular importance to our discussion is the (one) singlet in Fock
space spanned by the vacuum $|0\rangle$, w.r.t. which
operators in (\ref{1}) are normal ordered:
\begin{equation}
a_m|0\rangle=0;\quad a^\dagger_{-m}|0\rangle=0;\quad m>0\,.\label{2}
\end{equation}

Our strategy is to write a transition matrix element for a spin Hamiltonian
with source terms for single fermion operators as a path integral over
Grassmann variables. Consider then the Hamiltonian
\begin{equation}
H(t)=\alpha\cdot J+\sum_{m>0}(a_m^\dagger\sigma_m+a_{-m}\sigma_{-m}+
\sigma_{-m}^\dagger a^\dagger_{-m}+\sigma_{m}^\dagger a_{m})\,.\label{3}
\end{equation}
Here $\alpha\cdot J=\alpha_i J_i$ where $J_i$ are the Cartesian
components, $\alpha_i$ are real valued functions of time while
$\sigma_m\,,\sigma_m^\dagger$ are Grassmann valued functions of time
which anti-commute with the fermion creation and annihilation
operators.  The sources $\sigma_m\,,\sigma_m^\dagger$ eventually
facilitate construction of the bosonic equivalents of fermionic
tensor operators.  The dynamics is governed by the algebraic
(depending on generators only) Hamiltonian $H=\alpha\cdot J$.

We introduce the normalized fermionic coherent state [6]
\begin{equation}
%% FOLLOWING LINE CANNOT BE BROKEN BEFORE 80 CHAR
|\chi\rangle=\exp(-\sum_{m>0}(\chi_m^\dagger\chi_m+\chi_{-m}^\dagger\chi_{-m})/2+
\chi_m a^\dagger_m+\chi_{-m} a_{-m}))|0\rangle\,.\label{6}
\end{equation}
Here $\chi_m\,,\chi_m^\dagger$ are Grassmann variables which anti-commute with
the fermion creation and annihilation operators and $|0\rangle$ was defined in
(\ref{2}).  This state satisfies the normal completeness relation on Fock space
[6].

We consider a general transition matrix element,
$\langle\psi\,,t=1|\phi,t=0\rangle$,
between two arbitrary states in Fock space.  Inserting the identity at times
$t=0$ and $t=1$ in terms of the state (\ref{6}) and following the standard
procedure [6] we can write this matrix element as a path integral over the
coherent state (\ref{6}):
\begin{equation}
%% FOLLOWING LINE CANNOT BE BROKEN BEFORE 80 CHAR
\langle\psi\,,t=1|\phi,t=0\rangle\propto\int[d\chi^\dagger][d\chi]\langle\psi|\chi(1)\rangle
\langle\chi(0)|\phi\rangle e^{iS}\,,\label{7}
\end{equation}
where
\begin{equation}
S=\int_0^1dt[\chi^\dagger(i\frac{d}{dt}+\alpha)\chi+\chi^\dagger\sigma+
\sigma^\dagger\chi]\,.\label{8}
\end{equation}
Here we have introduced a number of short hand notations: $[d\chi^\dagger]
[d\chi]=\Pi_m [d\chi^\dagger_m][d\chi_m]$, $\chi^\dagger$ and $\sigma^\dagger$
are the row vectors $\chi^\dagger=(\chi_1\,,\chi_{-1}\,\ldots)$ and
$\sigma^\dagger=(\sigma_1\,,\sigma_{-1}\,\ldots)$, $\chi$ and $\sigma$ are
their adjoints, $\alpha=\alpha_i T_i$ with $T_i$ spanning the su(2) algebra.
Specifically the $T_i$ are the $2\Omega$ dimensional block diagonal matrices
given by the Kronecker product $I\otimes \sigma_i/2$ with $I$ the $\Omega$
dimensional identity matrix and $\sigma_i$ the Pauli spin matrix.

It should be pointed out that we integrate over $\chi(0)\,,\chi(1)$ in
(\ref{7}) with the boundary function
$\langle\psi|\chi(1)\rangle\langle\chi(0)|\phi\rangle$.  Note
that, apart from the exponential normalization of the coherent state, the
boundary function is a polynomial in the Grassmann variables.
Matrix elements of time ordered products of
fermionic operators follow from (\ref{7}) by differentiating w.r.t.
the $\sigma\,,\sigma^\dagger$ and setting them zero.

There is a subtlety connected to coherent state path integrals which has to be
kept in mind when deriving (\ref{8}).  The coherent state may be build on
different states and this may lead to ambiguities in the path integral.  We
have used a very specific coherent state (\ref{6}), build on the half-filled
shell, to derive the result (\ref{8}).  If other coherent states, such as the
one build on the fermion vacuum, are used an unwanted term is generated in the
action.  This term results from the normal ordering of the fermion operators
w.r.t. the vacuum on which the coherent state is build.  One can check directly
by calculating the trace (character) that this additional term is incorrect; it
violates the invariance of the trace under SU(2) transformations and leads to
an incorrect character.  A similar problem was also observed when deriving
(\ref{1a}) from SU(2) coherent states [7].

We now proceed to bosonize (\ref{7}).  In order to introduce bosonic
variables in the path integral, we insert the following identity
into (\ref{7}):
\begin{equation}
1=\int[dB]\exp(i\int_0^1dt\,\chi^\dagger B\chi)\delta(B)
=\int[dB][d\lambda]
\exp(i\int_0^1dt[\chi^\dagger B\chi+\frac{2}{\Omega}{\rm tr}(\lambda
B)])\,.\label{9}
\end{equation}
Here $B=B_i T_i\,,\lambda=\lambda_i T_i$ and we have used ${\rm
tr}(T_iT_j)=\Omega/2\delta_{ij}$.  A set of transformations is now performed
that decouples $\alpha$ from the fermionic degrees of freedom.  First we make
the transformations $B\rightarrow B-\alpha\,,\chi\rightarrow U\chi$ with
$U=\exp(i\theta_iT_i)\in$ SU(2).  This gives
\begin{equation}
%% FOLLOWING LINE CANNOT BE BROKEN BEFORE 80 CHAR
\langle\psi\,,t=1|\phi,t=0\rangle\propto\int[dB][d\lambda][d\chi^\dagger][d\chi]
\langle\psi|(U\chi)(1)\rangle
\langle(U\chi)(0)|\phi\rangle e^{iS}\,,\label{10}
\end{equation}
where
\begin{equation}
S=\int_0^1dt[\chi^\dagger(i\frac{d}{dt}+iU^\dagger\dot U+U^\dagger B U)\chi
+\frac{2}{\Omega}{\rm tr}(\lambda(B-\alpha)))+\chi^\dagger
U^\dagger\sigma+ \sigma^\dagger U\chi]\,.\label{11}
\end{equation}

We now make the change of integration variables $B=-i\dot U U^\dagger$ to
decouple the bosonic and fermionic degrees of freedom.  The
integration measure becomes $[dB]=J(U)[dU]$ with $[dU]$ the invariant group
measure and $J(U)$ the Jacobian associated with the change of variables.
Note that invariance of the group measure, $[d(U_0U)]=[dU]$, implies invariance
of the Jacobian $J(U_0 U)=J(U)$.  The Jacobian can therefore be
evaluated close to the identity and is given by $J[U]=\frac{\delta
B}{\delta\theta_i}|_{\theta_i=0}$.  A simple calculation shows that
$J[U]={\rm det}(\frac{d}{dt})$.  It is therefore a global normalization factor
which can be dropped from the path integral.

In order to obtain the desired form of the bosonic action we also introduce the
change of integration variables $\lambda=\lambda_3 h^\dagger T_3 h$ with
$\lambda_3\in[0,\infty)$ and $h\in$ U(1)$\setminus$SU(2).  The
integration
measure becomes $[d\lambda]=[dh] J(\lambda_3)[d\lambda_3]$ with $[dh]$ the
invariant measure on the coset and $J(\lambda_3)={\rm det}(\lambda_3^2)$.  With
these changes in variables we have
\begin{equation}
\langle\psi\,,t=1|\phi,t=0\rangle \propto  \int_0^\infty
J(\lambda_3)[d\lambda_3]\int
[dU][dh][d\chi^\dagger][d\chi]
\times\langle\psi|(U\chi)(1)\rangle
\langle (U\chi)(0)|\phi\rangle e^{iS}\,,\label{12}
\end{equation}
where
\begin{equation}
S=\int_0^1dt[i\chi^\dagger\dot\chi
+\frac{2\lambda_3}{\Omega}{\rm tr}(h^\dagger T_3 h(iU\dot
U^\dagger-\alpha)))+\chi^\dagger
U^\dagger\sigma+ \sigma^\dagger U\chi]\,.\label{13}
\end{equation}
The transformation $U\rightarrow h^\dagger U$ and the invariance
of $[dU]$ yields a decoupling of $U$ and $h$:
\begin{equation}
\langle\psi\,,t=1|\phi,t=0\rangle  \propto \int_0^\infty
J(\lambda_3)[d\lambda_3]\int
[dU][dh][d\chi^\dagger][d\chi]
\times\langle\psi|(h^\dagger U\chi)(1)\rangle
\langle (h^\dagger U\chi)(0)|\phi\rangle e^{iS}\,,\label{14}
\end{equation}
where
\begin{equation}
S=\int_0^1dt \!\![i\chi^\dagger\dot\chi
+\frac{2i\lambda_3}{\Omega}{\rm tr}(\dot U^\dagger T_3 U)
-\frac{2i\lambda_3}{\Omega}{\rm tr}(\dot h^\dagger T_3 h)
-\frac{2\lambda_3}{\Omega}{\rm tr}(T_3 h \alpha h^\dagger)
+\chi^\dagger U^\dagger h\sigma+ \sigma^\dagger h^\dagger U\chi]\,.\label{15}
\end{equation}
The final form is obtained by making the change of variables $U=\exp(i\theta
T_3)g$ with $g\in$U(1)$\setminus$SU(2).  The reason for doing this is that the
$\theta$ integration can be performed (see below).  This enforces the
constraint that the spin, $\lambda_3$, (see (\ref{1a})) is a constant
of motion.
The integration measure becomes $[dU]=[dg][d\theta]$ with $[dg]$ the invariant
measure on the coset.  This yields
\begin{eqnarray}
\langle\psi\,,t=1|\phi,t=0\rangle &\propto &\int_0^\infty
J(\lambda_3)[d\lambda_3]\int
[dg][d\theta][dh][d\chi^\dagger][d\chi]
\times\langle\psi|(h^\dagger\exp(i\theta
T_3)g\chi)(1)\rangle\nonumber\\
&&\times\langle (h^\dagger\exp(i\theta
T_3)g\chi)(0)|\phi\rangle
e^{i(S+\lambda_3(1)\theta(1)-
\lambda_3(0)\theta(0))}
\,,\label{16}
\end{eqnarray}
where
\begin{equation}
\begin{array}{ll}
S=\int_0^1dt&\!\![i\chi^\dagger\dot\chi
+\frac{2i\lambda_3}{\Omega}{\rm tr}(\dot g^\dagger T_3 g)
-\dot\lambda_3\theta
-\frac{2i\lambda_3}{\Omega}{\rm tr}(\dot h^\dagger T_3 h)\\
&-\frac{2\lambda_3}{\Omega}{\rm tr}(T_3 h \alpha h^\dagger)
+\chi^\dagger g^\dagger\exp(-i\theta T_3)
 h\sigma+ \sigma^\dagger h^\dagger\exp(i\theta T_3)g
\chi]\,.\label{17}
\end{array}
\end{equation}
The extra boundary term in (\ref{16}) has its origin in a partial integration
which was performed to shift the time derivative onto $\lambda_3$.

As before the matrix elements of time ordered products of fermionic operators
are evaluated by differentiating w.r.t. the $\sigma\,,\sigma^\dagger$ and
setting them zero.  This corresponds to making the insertions $\chi^\dagger
g^\dagger\exp(-i\theta T_3)h$ and $h^\dagger\exp(i\theta T_3)g\chi$ in the path
integral.

We first consider the case where no fermionic insertions are made, i.e, the
fermionic sources are set zero.  Doing this we note that the $\theta$ integral
can be explicitly performed to yield $\dot\lambda_3=0$, i.e., $\lambda_3$ is a
constant of motion.  This is no surprise since we note from (\ref{1a}) that
$\lambda_3$ is the spin or representation of SU(2).  Since the dynamics is
algebraic the SU(2) representation must be a constant of motion, which is
what we discovered above.  It is immediately clear that $\chi\,,\chi^\dagger$
and $g$ are spectators from a dynamical point of view.  Indeed the Hamiltonian
determining their dynamics is simply $H=0$.  The $\chi\,,\chi^\dagger$
and $g$ integrations yield, at most, a representation dependent normalization
factor.  The only relevant dynamical degree of freedom, which determines the
$\alpha$ dependency, is the $h$ field.

Since $\lambda_3$ is related to the SU(2) representation, we expect it to be
quantized in half-integer values.  Furthermore not all spin values are
admissible, but only those carried by fermion Fock space.  The way in which
this information is contained in the path integral is, as one may expect, on
the boundaries.  To see this note from the structure of the boundary function
$\langle\psi|(h^\dagger\exp(i\theta T_3)g)\chi(1)\rangle\langle
(h^\dagger\exp(i\theta
T_3)g)\chi(0)|\phi\rangle$ that each $\chi$ is multiplied by $\exp(\pm
i\theta/2)$.
Note, however, that the exponential normalization factor of the coherent state
is invariant under SU(2) transformations.  Keeping in mind that, apart from
this normalization factor, the boundary function is a polynomial in the
Grassmann variables, one notes from (\ref{16}) that the $\theta(0)\,,\theta(1)$
integrations fix the values of $\lambda_3$ on positive half-integer values
(recall $\lambda_3\in [0,\infty)$).  Since the $\chi$ are Grassmann valued,
only polynomials of a finite degree are admissible.  Hence not all spin values
are allowed, but only those consistent with the fermion statistics.  It is not
difficult to verify that all the SU(2) representations carried by fermion Fock
space occur with the correct multiplicities.  Furthermore the selection rule
that the initial and final states belong to the same SU(2)
representation is immediate.
{}From (\ref{17}) we finally infer for the SU(2) generators (currents)
the bosonization rule
\begin{equation}
\chi^\dagger T_i\chi\rightarrow -\frac{2\lambda_3}{\Omega}{\rm tr}(T_3 h T_i
h^\dagger)\,.\label{18}
\end{equation}
with the constant $\lambda_3$ determining the SU(2) representation.

Next we consider a single fermionic insertion at time $t_1\in(0,1)$.  Writing
the insertion out in component form results in a linear combination of two
terms, one containing $\exp(i\theta(t_1)/2)$ and the other containing
$\exp(-i\theta(t_1)/2)$.  Introducing a new variable $\lambda'_3=\lambda_3\pm
s(t-t_1)/2$ with $s(t)$ the step function, the $\theta$ integral can be
performed yielding $\dot\lambda'_3=0$.  This implies that $\lambda_3={\rm
const.}\pm s(t-t_1)/2$.  The integrations of $\theta$ on the boundary again
fixes the constant to be half-integer and imposes the selection rule that the
initial and final states belong to representations $j$ and $j\pm 1/2$,
respectively.  This illustrates the intertwining properties of the fermionic
operators which link representations $j$ to $j\pm 1/2$.

Once again it is clear that the $\chi^\dagger\,,\chi$ and $g$ are irrelevant
from a dynamical point of view.  These integrations yield a
normalization factor which only depends on the initial and final
representations, while the $\alpha$ dependency is determined by the $h$
integral.  The normalization factors only play a role in determining the
relative weight between the two terms in the above mentioned linear
combination.  It is thus actually more convenient to consider the
insertion of vertex operators [5], rather than fermionic operators which are
linear combinations of vertex operators with appropriate coupling coefficients.

A vertex operator [5] is the multiplet of operators $V_{j_2,m_2}[j_1,j_3]$
which maps the irrep $j_1$ to $j_3$.  As illustrated above, the
fermion operators are linear combinations of vertex operators intertwining from
the $j$ to the $j+1/2$ and $j-1/2$ representations, respectively, i.e.
of the operators $V_{1/2,m}[j,j\pm 1/2]$.  Their explicit form
can be read off from the fermionic insertions
$a^\dagger_m\,,a_{-m}\,,m>0$ which form a doublet under su(2):
\begin{eqnarray}
V_{1/2,1/2}[j,j-1/2]&=&\exp(-i\theta/2)h_{m,m}\,,\nonumber\\
V_{1/2,-1/2}[j,j-1/2]&=&\exp(-i\theta/2)h_{m,-m}\,,\nonumber\\
V_{1/2,1/2}[j,j+1/2]&=&\exp(i\theta/2)h_{-m,m}\,,\nonumber\\
V_{1/2,-1/2}[j,j+1/2]&=&\exp(i\theta/2)h_{-m,-m}\,.\label{19}
\end{eqnarray}
Note from the form of the $T_i$ matrices that the matrix elements of $h$ are
$m$ independent; hence no index $m$ appears on the vertex insertions.  The fact
that the vertex insertion is $m$-independent is merely a statement of the fact
that $a^\dagger_m\,,a_{-m}$ transform in the same way under SU(2) for all
$m>0$.  The different fermionic insertions are distinguished by the
$m$-dependent Grassmann variable which goes with the vertex insertion and also
takes care of the fermion statistics.  For $m<0$ one reads of vertex insertions
which are the complex conjugates of those above.  For SU(2) they are, however,
equivalent.  The above arguments can be generalized straightforwardly to more
insertions and more general vertex insertions can be read of in this way.

To calculate the character (trace) of $\alpha\cdot J$ in the present framework
we simply note that this corresponds to inserting the overlap of the coherent
state at time $t=0$ and $t=1$ as boundary function.  We do not impose the usual
periodic (anti-periodic) boundary conditions for commuting (anti-commuting)
variables.

To evaluate (\ref{16}) we have to resort to a particular parameterization of
the coset U(1)$\setminus$SU(2).  A convenient (non-global) parameterization is
\begin{equation}
h(z)=\exp(-z T_+)\exp(T_3\ln(1+z^*z))\exp(z^*T_-)\,.\label{21}
\end{equation}
which has the invariant measure
$dh(z)=1/(1+z^*z)^2dz^*dz$ on the coset [8].

A `free field realization' requires a transformation which allows
for the interpretation of the path integral as a path integral over the bosonic
coherent state $|z\rangle =\exp(-z^*z/2+z^*b^\dagger)|0\rangle$ (see [8]).  In
order to achieve this the transformation must change the invariant measure of
the path integral into the Euclidean measure, and also transform the kinetic
energy into a standard form.  In analogy with field theory [5] we call this a
free field realization.  One such transformation is $z^*\rightarrow
z^*\,,z\rightarrow z/(2\lambda_3-z^*z)$ and similarly for $w$.  Other
transformations are also possible, leading to different free field realizations
[9].  The end result is
\begin{eqnarray}
\langle\psi\,,t=1|\phi,t=0\rangle&\propto &\int_0^\infty
[d\lambda_3]\int
[dw^*][dw][dz^*][dz][d\theta][d\chi^\dagger][d\chi]\nonumber\\
&&\times\langle\psi|(h(z)^\dagger\exp(i\theta
T_3)g(w)\chi)(1)\rangle
\langle (h(z)^\dagger\exp(i\theta
T_3)g(w)\chi)(0)|\phi\rangle\nonumber\\
&&\times e^{i(S+\lambda_3(1)\theta(1)-
\lambda_3(0)\theta(0))}
\,,\label{23}
\end{eqnarray}
where
\begin{equation}
\begin{array}{ll}
S=\int_0^1dt&\!\![i\chi^\dagger\dot\chi
+iw\dot w^*+i\lambda_3\frac{d}{dt}\ln(2\lambda_3-w^*w)
-\dot\lambda_3\theta
-iz\dot z^*-\\
&i\lambda_3\frac{d}{dt}\ln(2\lambda_3-z^*z)
+\alpha_+(2\lambda_3-z^*z)
+\alpha_0z^*(z^*z-\lambda_3)
+\alpha_-z\\
&+\chi^\dagger g(w)^\dagger\exp(-i\theta T_3)
 h(z)\sigma+ \sigma^\dagger h(z)^\dagger\exp(i\theta T_3)g(w)
\chi]\,.\label{24}
\end{array}
\end{equation}
In terms of the free fields we infer the bosonization rules
(see (\ref{18}))
\begin{eqnarray}
\chi^\dagger T_+\chi&\rightarrow & z^*(2\lambda_3-z^*z),\nonumber\\
\chi^\dagger T_0\chi&\rightarrow & (z^*z-\lambda_3),\nonumber\\
\chi^\dagger T_-\chi&\rightarrow & z\,.\label{25}
\end{eqnarray}

When no vertex insertions are made the $\theta$ integral can be performed and
the term $\lambda_3\frac{d}{dt}\ln(2\lambda_3-z^*z)$ is a total time
derivative.  However, when vertex insertions are made, it is no longer a total
time derivative.  The $\theta$ integral can, however, still be
evaluated using step
functions as outlined before.  The result is that this term leads to a total
time derivative as well as insertions of the form $(2\lambda_3-z^*z)^{\pm 1/2}$
at the same time as the vertex insertion.  This has to be taken into account
when the vertex insertions are determined in the free field realization.  The
result is (see (\ref{19}))
\begin{eqnarray}
V_{1/2,1/2}[j,j-1/2]&=&\exp(-i\theta/2)(2\lambda_3-z^*z)\,,\nonumber\\
V_{1/2,-1/2}[j,j-1/2]&=&-\exp(-i\theta/2)z\,,\nonumber\\
V_{1/2,1/2}[j,j+1/2]&=&\exp(i\theta/2)z^*\,,\nonumber\\
V_{1/2,-1/2}[j,j+1/2]&=&\exp(i\theta/2)\,.\label{26}
\end{eqnarray}

It is informative to write down the bosonization rules (\ref{25}) and
vertex operators (\ref{26}) on the second quantized level (for the
purposes of evaluating in the path integral they are, of course, not
required).  Noting that $\theta$ is the canonical conjugate momentum
of $\lambda_3$ we introduce the operators $\theta\rightarrow
q\,,\lambda_3\rightarrow p$ where $p$ and $q$ satisfy canonical
commutation relations $[q,p]=i$.  Interpreting (\ref{23}) as a path
integral over the bosonic coherent state we immediately infer
$f(z^*,z)\rightarrow :f(b^\dagger,b):$ where : : denotes normal
ordering.  Note that we have been led quite naturally to introduce a
complex of boson Fock spaces [5].

{}From (\ref{25}) we obtain the well known boson representation (Dyson mapping
[4])
%\begin{eqnarray}
%J_+&\rightarrow& b^\dagger(2p-b^\dagger b),\nonumber\\
%J_0&\rightarrow& (b^\dagger b-p),\nonumber\\
%J_-&\rightarrow& b\,.\label{27}
%\end{eqnarray}
\begin{equation}
J_+\rightarrow b^\dagger(2p-b^\dagger b),\quad
J_0\rightarrow (b^\dagger b-p),\quad
J_-\rightarrow b\,.\label{27}
\end{equation}
and from (\ref{26}) we obtain for the vertex operators
\begin{eqnarray}
V_{1/2,1/2}[j,j-1/2] &=& \exp(-iq/2)(2p-b^\dagger b)\,,\\
V_{1/2,-1/2}[j,j-1/2] &=& -\exp(-iq/2)b\,,\\
V_{1/2,1/2}[j,j+1/2] &=& \exp(iq/2)b^\dagger\,,\\
V_{1/2,-1/2}[j,j+1/2] &=& \exp(iq/2)\,.\label{28}
\end{eqnarray}
It is easy to check that these operators have the correct tensorial properties
under (\ref{26}).

It is well known that the free field realization of SU(2) is reducible
[5], the finite irreducible representation being carried by the
subspace with $2j$ or fewer bosons (physical subspace).  The
equivalent of a transition matrix element on the fermionic level is
therefore a transition matrix element between two physical states.
The way in which this information is embodied in the path integral
(\ref{23}) is via the boundary function which explicitly constructs
the bosonic equivalent of a fermionic state.  No additional input is
required.  When no vertex insertions are made, it is obvious that only
physical states propagate internally since the physical subspace is
invariant under the SU(2) generators.  When vertex insertions are made
one has to ensure that this remains the case, i.e., the vertex
insertions should not introduce unphysical states.  The vertex
insertions constructed in (\ref{28}) clearly have this property, as is
to be expected since they were derived from fermionic insertions which
are physical.  Once again no additional input is required.

In conclusion we have shown that a transition matrix element in
fermion Fock space for a spin Hamiltonian, expressed as a path
integral over Grassmann variables, can be expressed as a path integral
over complex variables (bosonization).  We have identified the bosonic
equivalents of the SU(2) generators and of the single fermion
operators expressed as vertex operators.  This allows the construction
of the free field realization (Dyson boson mapping) of the SU(2)
generators and vertex
operators, and the explicit demonstration that the vertex operators
only intertwine between physical subspaces.

This research was supported by grants from the Foundation for Research
Development.

\end{document}